\documentclass[]{aastex61}
\usepackage{CJK}

\newcommand\aastex{AAS\TeX}

\newcommand\cmminussquared{cm$^{-2}$}
\newcommand\cmmt{cm$^{-2}$}
\newcommand\cmmth{cm$^{-3}$}
\newcommand\um{$\mu$m}
\newcommand\mjybm{mJy beam$^{-1}$}
\newcommand\jybm{Jy beam$^{-1}$}
\newcommand\nthp{N$_2$H$^+$}
\newcommand\hcop{HCO$^+$}
\newcommand\hcn{HCN}

\received{...}
\revised{...}
\accepted{...}
\submitjournal{ApJ}

\shorttitle{\aastex\ BISTRO: Serpens Main}
\shortauthors{Kwon et al.}
\begin{document}
\begin{CJK*}{UTF8}{gbsn}

\title{B-fields in Star-Forming Region Observations (BISTRO): 
Magnetic Fields in the Filamentary Structures of Serpens Main}

\correspondingauthor{Woojin Kwon}
\email{wkwon@snu.ac.kr}

\AuthorCallLimit=200

\author[0000-0003-4022-4132]{Woojin Kwon}
\affiliation{Department of Earth Science Education, Seoul National University, 1 Gwanak-ro, Gwanak-gu, Seoul 08826, Republic of Korea}
\affiliation{SNU Astronomy Research Center, Seoul National University, 1 Gwanak-ro, Gwanak-gu, Seoul 08826, Republic of Korea}

\author[0000-0002-8557-3582]{Kate Pattle}
\affiliation{Department of Physics and Astronomy, University College London, Gower Street, London WC1E 6BT, United Kingdom}
\affiliation{Centre for Astronomy, School of Physics, National University of Ireland Galway, University Road, Galway H91 TK33, Ireland}

\author{Sarah Sadavoy}
\affiliation{Department for Physics, Engineering Physics and Astrophysics, Queen's University, Kingston, ON, K7L 3N6, Canada}

\author[0000-0002-8975-7573]{Charles L. H. Hull}
\affiliation{National Astronomical Observatory of Japan, Alonso de C\'ordova 3788, Office 61B, Vitacura, Santiago, Chile}
\affiliation{Joint ALMA Observatory, Alonso de C\'ordova 3107, Vitacura, Santiago, Chile}
\affiliation{NAOJ Fellow}

\author[0000-0002-6773-459X]{Doug Johnstone}
\affiliation{NRC Herzberg Astronomy and Astrophysics, 5071 West Saanich Road, Victoria, BC V9E 2E7, Canada}
\affiliation{Department of Physics and Astronomy, University of Victoria, Victoria, BC V8W 2Y2, Canada}

\author[0000-0003-1140-2761]{Derek Ward-Thompson}
\affiliation{Jeremiah Horrocks Institute, University of Central Lancashire, Preston PR1 2HE, UK}

\author[0000-0002-9289-2450]{James Di Francesco}
\affiliation{NRC Herzberg Astronomy and Astrophysics, 5071 West Saanich Road, Victoria, BC V9E 2E7, Canada}
\affiliation{Department of Physics and Astronomy, University of Victoria, Victoria, BC V8W 2Y2, Canada}

\author[0000-0003-2777-5861]{Patrick M. Koch}
\affiliation{Academia Sinica Institute of Astronomy and Astrophysics, No.1, Sec. 4., Roosevelt Road, Taipei 10617, Taiwan}

\author{Ray Furuya}
\affiliation{Tokushima University, Minami Jousanajima-machi 1-1, Tokushima 770-8502, Japan}
\affiliation{Institute of Liberal Arts and Sciences Tokushima University, Minami Jousanajima-machi 1-1, Tokushima 770-8502, Japan}

\author[0000-0001-8746-6548]{Yasuo Doi}
\affiliation{Department of Earth Science and Astronomy, Graduate School of Arts and Sciences, The University of Tokyo, 3-8-1 Komaba, Meguro, Tokyo 153-8902, Japan}

% ...



\begin{abstract}

We present 850 \um\ polarimetric observations toward the Serpens Main
molecular cloud {obtained} using the POL-2 polarimeter on the James Clerk
Maxwell Telescope (JCMT) as part of the B-fields In STar-forming
Region Observations (BISTRO) {survey}.  These observations probe
the magnetic field morphology of the
Serpens Main molecular cloud on about 6000 au scales, which consists
of cores and six filaments with different physical properties such as
density and star formation activity. {{Using} the histogram of relative orientation (HRO) technique, we} find that magnetic fields
are parallel to filaments in less dense filamentary structures {where} $N_{H_2}
< 0.93\times 10^{22}$ \cmminussquared\ {(magnetic fields
perpendicular to density gradients)}, while being perpendicular
to {filaments (magnetic fields parallel to density gradients)} 
in dense filamentary structures with star {formation} activity.
{Moreover, applying the HRO technique to denser core {regions},
we find that magnetic field orientations change
{to become} perpendicular to density gradients again
at $N_{H_2} \approx 4.6 \times
10^{22}$ \cmminussquared. This can be interpreted as {a signature of} core formation. 
At $N_{H_2} \approx 16 \times 10^{22}$ \cmminussquared\
magnetic fields change back {to being} parallel to density gradients {once again}, 
which {can be understood to be due to} magnetic fields being dragged in {by infalling material}.}
In addition, we estimate the magnetic field strengths
of the filaments ($B_{POS} = 60-300~\mu$G))
{using} the Davis-Chandrasekhar-Fermi method and discuss 
whether the filaments are gravitationally unstable based {on}
magnetic field and turbulence energy densities.

\end{abstract}

\keywords{
Interstellar medium (847), 
Interstellar magnetic fields (845), 
Interstellar filaments (842),
Molecular clouds (1072), 
Star formation (1569), 
Polarimetry (1278), Submillimeter astronomy (1647)}



\section{Introduction} \label{sec:intro}

Magnetic fields are thought to play a key role in various stages
of star formation over a large range of scales.  {For example,
ordered magnetic fields have been
detected in molecular clouds on 0.1 pc and larger
scales} \citep[e.g.,][]{2013A&A...550A..38P,2014arXiv1404.2024L},
{suggesting} that magnetic fields play a role in forming such
structures.  {However}, interferometric observations have
shown that {the} magnetic fields of young stellar objects (YSOs) {vary}
in morphology on about 100 au and smaller scales.  {Surveys of large samples of YSOs} have reported that magnetic fields are randomly
oriented {with respect to bipolar outflows}
\citep[e.g.,][]{{2013ApJ...768..159H},{2014ApJS..213...13H}}, while
hourglass shaped fields have been reported {toward a few of the youngest
protostellar systems} \citep[e.g.,][]{{2006Sci...313..812G},
{2013ApJ...769L..15S}}.  Recent polarimetric observations {made using} the
Atacama Large Millimeter/submillimeter Array (ALMA) have been
{revealing} various detailed {polarization and magnetic field structures}
on {scales of 100 au and smaller} toward many YSOs, with high
angular resolution and sensitivity{ including} a clear hourglass morphology
\citep[e.g.,][]{{2018MNRAS.tmp..552M},{2019ApJ...879...25K}}, {magnetic field structures tracing} turbulence {and/or outflow cavities} \citep[e.g.,][]{{2017ApJ...847...92H},{2018ApJ...869..115S},
{2019ApJ...885..106L}}, and {polarization patterns} {due to scattering {of mm/submm radiation}} by large grains
\citep[e.g.,][]{{2018ApJ...855...92C},{2018ApJ...854...56L},
{2019ApJS..245....2S}}.

The B-fields In STar forming Region Observations (BISTRO) {survey}
\citep{{2017ApJ...842...66W}} has been studying 
{the role of magnetic fields} in star formation at submillimeter wavelengths {on about 2000--7000 au 
scales}, 
using the James Clerk Maxwell Telescope (JCMT) and
its polarimeter POL-2.
This project {began} 
polarimetric observations in 2016 and {was} completed in
early 2019, {targeting 16} fields {in} nearby 
star forming regions: e.g., 
Orion A \citep{{2017ApJ...842...66W},{2017ApJ...846..122P}}, 
Ophiuchus \citep{{2018ApJ...859....4K},{2018ApJ...861...65S},{2019ApJ...877...43L},{2019ApJ...880...27P}}, 
IC 5146 \citep{{2019ApJ...876...42W}}, 
Barnard 1 \citep{{2019ApJ...877...88C}},
NGC 1333 \citep{{2020ApJ...899...28D}}, and
Auriga \citep{{2020arXiv201204297B}}.
In addition to {studies of individual regions}, 
\citet{Yen:2020tj} found {{using} all of the BISTRO data} that magnetic fields are misaligned with outflows by
about $50\degr$ on 0.05-0.5 pc scales.
The BISTRO project was extended {to include more distant
and more massive star-forming
regions \citep[BISTRO-2;][]{{Pattle:2018eb},{2020arXiv201213060A}}
and to {observe a wider range of} evolutionary stages} and {star formation environments}
(BISTRO-3).

As part of {the} BISTRO {survey}, this paper focuses on the Serpens Main molecular
cloud, which is located at a distance of 415 pc
\citep{{2010ApJ...718..610D},{Herczeg:2019cp}}.  Serpens Main consists of
two subclusters--- North-West (hereafter NW) and South-East {(SE)}
--- and has largely been studied as an active, {intermediate-mass} 
{star-forming} region.  For example, several tens of YSOs have 
been detected in the region by the Spitzer Space
{Telescope} \citep{{2007ApJ...663.1149H},{2009ApJS..181..321E}},
and it was found that younger YSOs (Class 0/I) are located in the
central regions of the subclusters, while older YSOs (Class II/III)
are {{more} widely distributed, and preferentially located} in the SE subcluster
\citep{{2007ApJ...669..493W}}.  {Furthermore}, the SE
subcluster shows larger velocity gradients compared to the NW {subcluster}
\citep{{2010MNRAS.409.1412G}}.  \citet{{2014ApJ...797...76L}} mapped the molecular
cloud {at} $7\arcsec$ resolution in the 3 mm continuum and the
\nthp, \hcop, and \hcn\ {(1-0)} molecular lines, using the Combined Array
for Research in Millimeter-wave Astronomy (CARMA).  They identified six
filamentary structures {using} the \nthp\ (1-0) channel maps.  Interestingly,
two filaments are less dense and do not have star formation activity,
while the others are dense enough for gravitational collapse 
and have YSOs.
The {properties of, and star formation activity within, these filaments} need to
be {investigated} alongside {their} magnetic fields, {and} {are} the main {topics} of this
paper.  {Previously}, \citet{2009ApJS..182..143M} detected linear
polarizations toward Serpens Main {using the} JCMT and SCUPOL {(the predecessor to POL-2)}, but
the polarized signal was detected only in {limited regions in the centers}
of the two subclusters, which {made} it impossible to study the magnetic
fields in the filamentary structures.
{Conversely}, \citet{2010ApJ...716..299S} {have} studied the
large-scale magnetic fields around the molecular cloud 
using near-infrared imaging polarimetry. The
detailed magnetic fields within the molecular cloud, however, could not be
{studied using such data}.

In this paper, we describe how the data have been taken
and reduced in Section \ref{sec:obs}. {Detected polarization measurements} and inferred
magnetic fields are introduced in Section \ref{sec:results}, followed by
discussions on the relationships between the physical properties of {the}
filaments, including {their} magnetic fields and star formation activities
(Section \ref{sec:discuss}). {We} address {the} column densities at which
the relative directions of magnetic fields {{and column density structures} change},
in Section \ref{sec:discuss}. 
Finally, we summarize {our} results {in Section \ref{sec:final}}.

\section{Observations} \label{sec:obs}

We observed the Serpens Main molecular cloud using {the 15m-diameter} JCMT, 
as part of {the} BISTRO {survey}.  The {POL-2 polarimeter} was {inserted into} the ray pathway
{of} the Submillimeter Common-User Bolometer Array 2 
\citep[SCUBA-2;][]{{Holland:2013eq}} {camera, in order to make linear polarization measurements}. Both 450 \um\ and 850 \um\
{measurements are taken} simultaneously, but only the 850 \um\ data, 
with an angular resolution of $14\arcsec$, are
{used} in this study.
POL-2 observations adopt a modified version of the standard SCUBA-2 DAISY
pattern, scanning the sky at a speed of $8 ''\textrm{ s}^{-1}$ with
a {half-wave plate rotation speed of} 2 Hz.  This mode allows
{each} $4''$ pixel to {contain samples from a full} revolution of the half-wave plate
\citep{2016SPIE.9914E..03F}, and provides 13 arcminute-diameter mapping coverage with a
central 3 arcminute-diameter area of the best sensitivity.  
Two pointings have been observed, each
for 14 hours (21 {repeats of} 40 minutes long {observing blocks}) in
{JCMT} weather band 2 ($0.05 < \tau < 0.08$ at 225 GHz), {in order}
to cover the whole Serpens
Main molecular cloud {to good} sensitivity.  {The two pointings were centered on} (R.A., {decl.}) =
(18:29:59.370, +01:12:52.00) and (18:29:49.530, +01:16:11.10) in
J2000 {coordinates} for the southeast (SE) and the northwest (NW) fields, respectively.  
The SE field {was observed} in 2016 April to May, while the NW field {was}
observed in 2017 March to May.

{Standard POL-2} data reduction procedures 
{were} followed {using} the 
{$pol2map$\footnote{http://starlink.eao.hawaii.edu/docs/sc22.htx/sc22.html} routine}, {which has}
recently {been} added to the {\it Starlink} software suite \citep{{2014ASPC..485..391C}}.
The {$pol2map$ routine} makes {Stokes I (total intensity)}, Stokes Q, and Stokes U maps
from {POL-2} time-stream data.  At the final stage of making images {using $pol2map$}, the
$skyloop$\footnote{http://starlink.eao.hawaii.edu/docs/sun258.htx/sun258ss72.html}
algorithm, an implementation of the SCUBA-2 iterative mapmaking
algorithm $makemap$ \citep{{2013MNRAS.430.2545C}} which improves image
consistency by reducing multiple observations simultaneously, was
{used.  We} applied the `August 2019' {instrumental polarization (IP)} model\footnote{{https://www.eaobservatory.org/jcmt/2019/08/new-ip-models-for-pol2-data/}} to the data, with
our final total intensity map serving as the model for {estimating} artificial
polarized emission to be corrected for.  In addition to the {typical POL-2}
procedures \citep[e.g.,][]{{2020arXiv201109765P}}, {additional} steps
{were taken to combine the SE and NW
fields}.  Data from the two pointings were initially reduced
separately to make initial total intensity maps. These maps were
combined to {produce} an initial coadded total intensity map, which was
used to generate a signal-to-noise-based mosaic mask for use in the
next step. This mask constrains the mapmaker by defining a fixed
area containing astrophysical signal \citep{{2013MNRAS.430.2545C}}.
The data from each field {were} again reduced separately, at this stage
invoking the {$skyloop$} algorithm and using the mask defined based
on the initial coadded $I$ map, to generate Stokes $Q$ and $U$ maps,
and an improved Stokes $I$ map. The $Q$ and $U$ maps were corrected
for instrumental polarization using the improved Stokes $I$ map and
the August 2019 IP model.  Finally, the maps of the two fields were
coadded, to produce a final mosaicked set of Stokes $I$, $Q$, and
$U$ maps from which polarization fractions and directions {were}
calculated {using} the equations: $p = PI/I$ and $\psi = 0.5
\tan^{-1}(U/Q)$, where $p$ is polarization fraction and $PI$ is
polarized intensity.  When calculating the polarization intensities,
classical de-biasing was adopted, {such that} $PI = \sqrt{Q^2 + U^2 -
\sigma^2}$, where $\sigma^2 = (Q^2 \sigma_Q^2 + U^2 \sigma_U^2)/(Q^2
+ U^2)$. {{Finally,}} the {raw POL-2} power unit of pW {was} converted to
{intensity units} of Jy beam$^{-1}$ {using the POL-2 flux conversion factor (FCF) of 725 Jy beam$^{-1}$ pW$^{-1}$, determined} by multiplying the SCUBA-2 850
\um\ {FCF} \citep[537 Jy beam$^{-1}$
pW$^{-1}$;][]{{2013MNRAS.430.2534D}} by a factor 1.35 to account
for additional losses in POL-2 at 850 \um\ \citep{{2016SPIE.9914E..03F}}.
All the maps {were made {using}}
the default $4\arcsec$ pixels, and the {noise level achieved in the final} total intensity map is about 4 \mjybm.

\section{Results} \label{sec:results}

\begin{figure}
\includegraphics[scale=0.30]{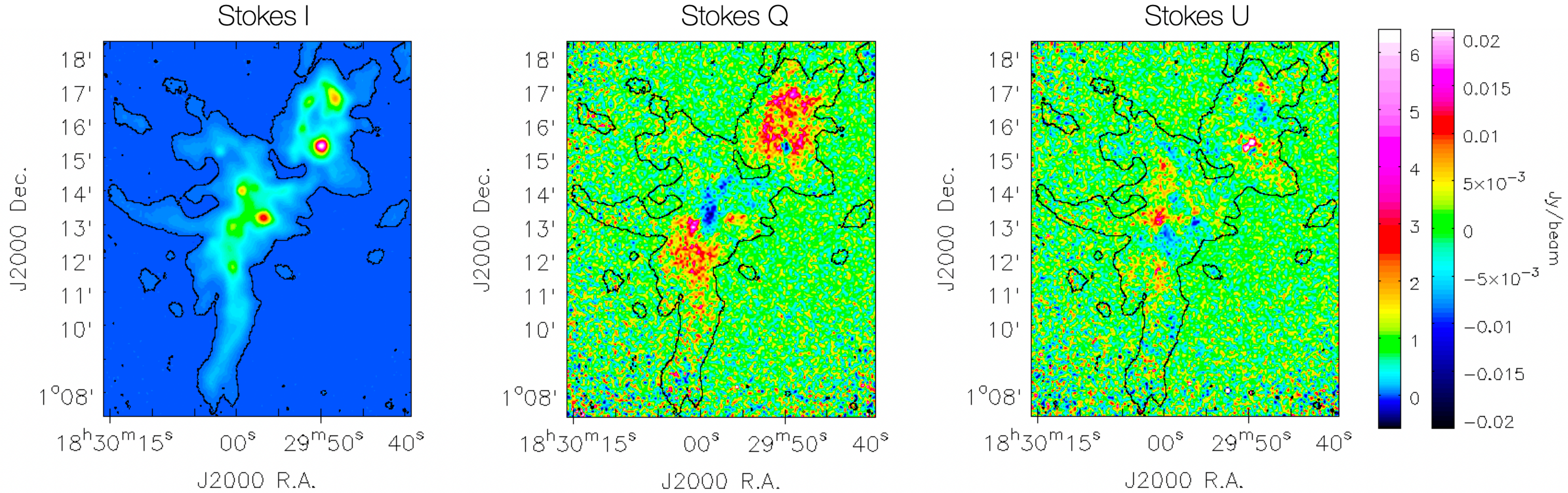}
\caption{Stokes I, Q, and U maps {(left, center, and right, respectively)} toward the
Serpens Main molecular cloud obtained {at 850 \um\ } {using} JCMT/POL-2
{as part of the BISTRO survey}. The contours indicate 12 \mjybm\ in the total intensity map, which is 
{approximately} the $3\sigma$ level.
\label{fig_serp}}
\end{figure}

Figure \ref{fig_serp} shows {our} Stokes I,
Q, and U maps, {demonstrating} linear polarization
{signal} toward Serpens Main.
The {noise levels achieved} are 3.8, 3.1, and 2.9 \mjybm\, respectively. 
As shown {in the figure}, linear polarizations have {been detected well} over a
large region, not only {in} the {centers of the} two subclusters. Previous
JCMT/SCUPOL observations also detected clear linear {polarization signal},
but {these {detections} were limited to only} the central regions
\citep{2009ApJS..182..143M} {due to lack of sensitivity}.

\begin{figure}
\plotone{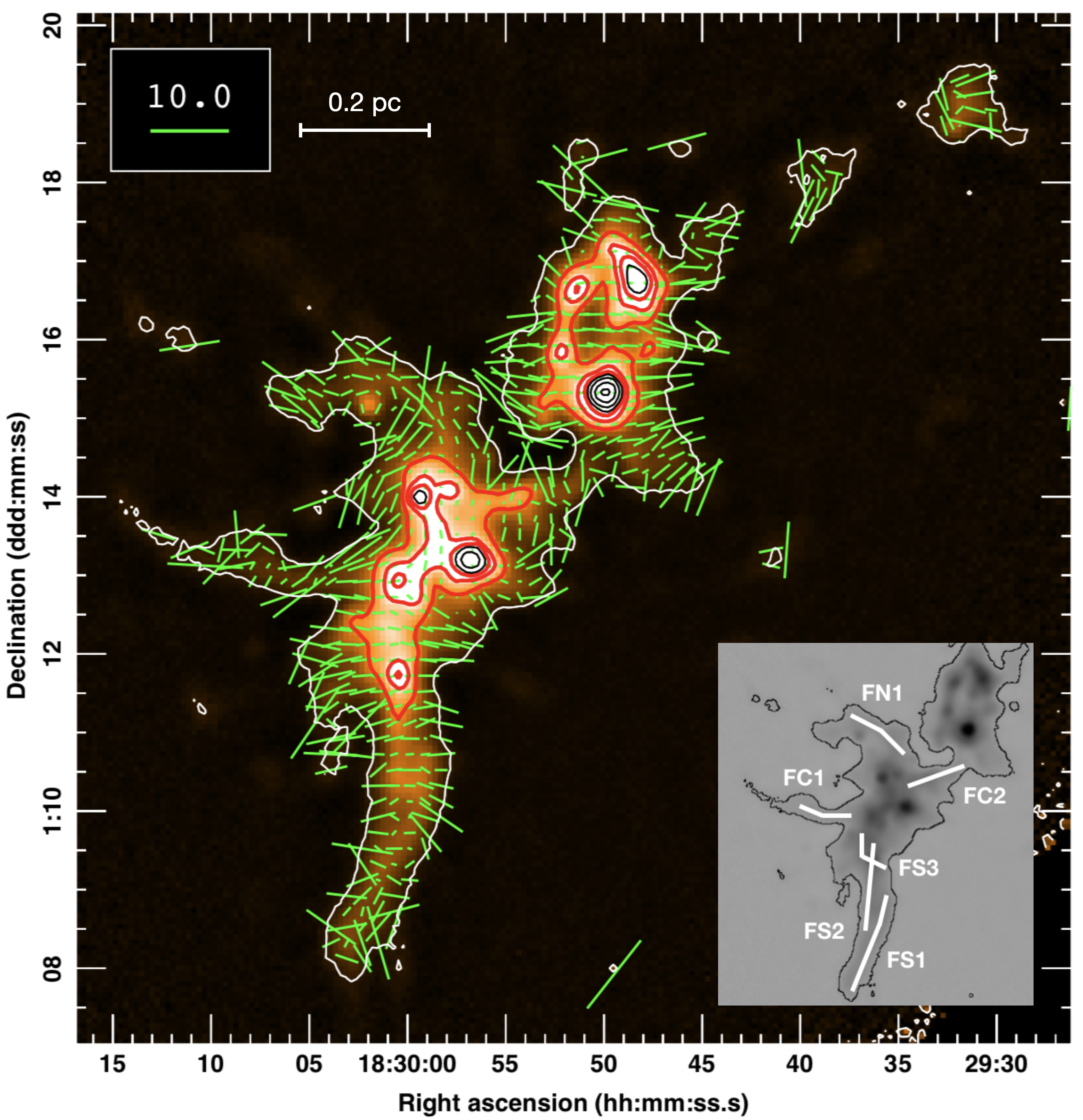}
\caption{Serpens Main magnetic field {directions (green segments)}
inferred from {850 \um} JCMT POL-2 observations by {rotating polarization directions by} $90\degr$. The green segment {lengths} indicate
polarization fractions; {{a polarization fraction of 10\% is shown} by the bar}
at the top left.  {The white bar indicates a length scale
at {415 pc, the distance of Serpens Main}.} 
The color {scale} and contours
show {the} total intensity {distribution}.  Contour levels are 0.042,
0.362, 0.725, 1.090, 1.450, 2.170, 3.620, and 5.800 \jybm.  The
inset {{panel} identifies} {the} six filaments {discussed in the text}.
\label{fig_serppol}}
\end{figure}

The Serpens Main molecular cloud has six filamentary structures
{located} {predominantly} in the SE subcluster, as shown in the inset {panel}
of Figure \ref{fig_serppol}.
\citet{{2014ApJ...797...76L}} observed Serpens Main in the
3 mm continuum and three {J=1-0} molecular lines of \nthp, \hcop, and
\hcn\ using CARMA at $7 \arcsec$ resolution.
They analyzed its structures and kinematics {using} the
dendrogram technique \citep{{Rosolowsky:2008ho}}. In particular,
they identified six filamentary structures based on the integrated
intensity and channel maps of \nthp\ (1-0), which is known to trace high
density regions: {e.g., $n_{crit}$(\nthp\ 1-0) = $6.1\times10^4~cm^{-3}$
at T = 10 K \citep{{Shirley:2015et}}}.
Some properties of the filaments selected from 
\citet{{2014ApJ...797...76L}} are listed in Table \ref{tab_filaments},
{and refer to {the filaments identified in the inset panel of Figure \ref{fig_serppol}}.}
Note that there are two types of filaments: FN1, FC1, and FC2 have no
YSOs, while the other three filaments contain YSOs. Furthermore,
{filaments} FN1 and FC1 are less dense and have large
velocity gradients, while the other filaments are
denser and have {relatively small velocity gradients}. Finally,
{the} linear densities of {the} FN1 and FC1 filaments are comparable to or lower
than {the} critical values {for} isothermal, self-gravitating
cylinders without magnetic support, {above which they} can gravitationally contract
\citep{{1964ApJ...140.1056O},{2014ApJ...797...76L}}.

\begin{deluxetable*}{ccccccccccc}
\tablecaption{Selected physical properties of filaments \label{tab_filaments}}
\tablehead{
\colhead{Filament} & 
\colhead{Width} &
\colhead{$\nabla V$} & \colhead{$\langle \sigma \rangle$} &
\colhead{$\langle \sigma \rangle _{nt}$} & \colhead{N$_{H_2}$} &
\colhead{Mass} & \colhead{M$_L$} & \colhead{M$_{L,crit}$} & \colhead{YSOs}\\
 & 
(pc) &
(km s$^{-1}$ pc$^{-1}$) & (km s$^{-1}$) & (km s$^{-1}$ &)
 ($\times 10^{22}$ cm$^{-2}$) & ($M_\sun$) & ($M_\sun$ pc$^{-1}$) & ($M_\sun$ pc$^{-1}$)}
\startdata
FN1 & 0.03 & $3.2 \pm 0.08$ & 0.27 & 0.21 & 3.6 &  3.79 & 18.0 & 23.7 & no \\
FC1 & 0.04 & $4.8 \pm 0.14$ & 0.25 & 0.24 & 3.5 &  4.76 & 28.0 & 20.0 & no \\
FC2 & 0.05 & $0.7 \pm 0.07$ & 0.21 & 0.20 & 5.2 & 10.57 & 52.9 & 22.4 & no \\
FS1 & 0.04 & $0.8 \pm 0.03$ & 0.13 & 0.12 & 4.3 & 16.33 & 49.5 & 19.2 & yes \\
FS2 & 0.03 & $1.5 \pm 0.08$ & 0.24 & 0.23 & 7.9 & 17.95 & 74.8 & 20.2 & yes \\
FS3 & 0.03 & $0.8 \pm 0.07$ & 0.23 & 0.22 & 9.3 & 14.08 & 82.8 & 20.5 & yes \\
\enddata
\tablecomments{All the values and information {in this table} come from 
\citet{{2014ApJ...797...76L}}, {who} identified the filaments
{using CARMA} \nthp\ (1-0) observations with {an angular resolution of} $7\arcsec$.
{The columns {list} filament widths, velocity gradients along the filaments ($\nabla V$), 
velocity dispersions ($\langle \sigma \rangle$), non-thermal velocity dispersions
($\langle \sigma \rangle_{nt}$), column densities of molecular hydrogen (N$_{H_2}$),
filament masses, filament masses per length (M$_L$), critical filament masses per length
(M$_{L,crit}$), and whether YSOs are found.}}
\end{deluxetable*}

\subsection{Field morphology} \label{subsec:tables}

Magnetic fields, {and specifically} the field morphology projected on the
plane of sky, can be studied {using} polarized thermal emission from dust {grains}.
Non-spherical dust grains {spin around} their minor axes
due to radiative anisotropic {torques}. Within a magnetic field,
such spinning grains {will have their minor axes aligned} parallel to
the field \citep[e.g.,][]{{2007MNRAS.378..910L}}. Therefore, the thermal
emission of non-spherical dust grains is linearly polarized
perpendicular to the field. 
This allows us to infer magnetic field directions by {rotating linear polarisation angle measurements by} $90\degr$.
Note that {{in contrast, at} short wavelengths {such as} those in the optical regime,
polarization is mainly caused by preferential absorption {of unpolarised background emission by}
aligned grains, rather than by the emission from grains.}
Thus, at those shorter wavelengths, {linear
polarization and magnetic field directions} are parallel to {one another}.

Figure \ref{fig_serppol} shows magnetic field directions 
{projected onto the plane of sky}
across Serpens Main {as} green
segments, {inferred by rotating the linear polarization angles by 90\degr},
overlaid {on the 850\um} total intensity map.
{Data} points with a signal-to-noise ratio (SNR) {in} total intensity {of} 
greater than 30 and a polarization fraction smaller than 15\% have been
selected after {binning up by a factor of 3 in pixel width}.
{{This}} binning
{yields} a {pixel size of $12\arcsec$,} comparable to the angular resolution
of $\sim14\arcsec$.
The lengths of the segments indicate polarization fractions,
which are relatively large (up to about 10\%) in the {low-intensity}
boundary regions, and get smaller going into the dense core regions.
The polarization fractions are {discussed quantitatively} in the next
subsection.

\begin{figure}
\plotone{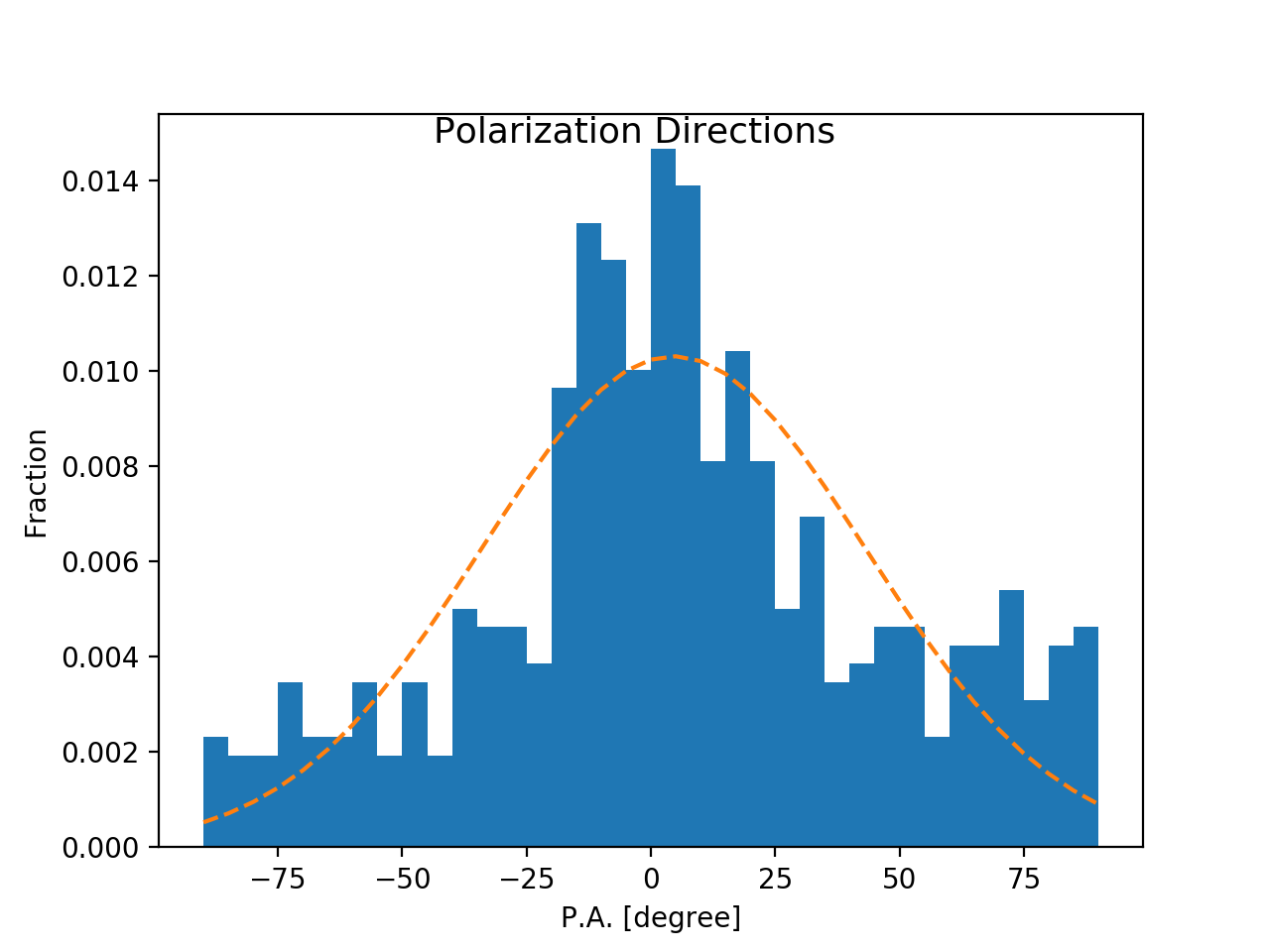}
\caption{{A} histogram of polarization {position angles (P.A.)} in Serpens Main.
The orange dashed line {shows} a Gaussian fit
to the histogram, {with} $\bar{\theta} = 4.5\degr$ and $\sigma = 38.7\degr$.
\label{fig_histoall}}
\end{figure}

The {distribution of} {polarization directions} is shown as a
histogram in Figure \ref{fig_histoall}. The most common direction is
P.A. = $4.5\degr$, suggesting that 
an east-west magnetic field dominates.
This direction is {somewhat} consistent with that {measured by {\it Planck}}
toward Serpens Main.
We estimate an average direction of $P.A. = -20 \degr$ from the
{\it Planck} polarimetric data at 353 GHz \citep{{2016A&A...586A.138P}}, 
assuming a 10 arc-minute
beam at the {J2000} center position of Serpens Main, (R.A., decl.) = (18:29:55,
+01:13:00).
Since the histogram {shown in} Figure \ref{fig_histoall}
is a distribution of {polarization angles measured on smaller scales} than those of
the {\it Planck} data, the
{means} of the directions {are} not expected to agree entirely, {and the} difference of $\sim25\degr$ can be understood.
The polarization direction of {\it Planck} data {matches the} 
{direction of elongation} of the Serpens
Main molecular cloud, {suggesting} that the large-scale 
magnetic field is perpendicular to the {overall elongation of the molecular gas} structures.
This large-scale field has also been reported {in} near-infrared polarimetric
observations sensitive to the {periphery} of the molecular cloud
\citep{{2010ApJ...716..299S}}.

The magnetic field directions {within} individual filaments with {differing}
physical properties {are} one of the key focuses {of} this study.
Figure \ref{fig_serppol} {shows that} the southern filaments, 
{FS1, FS2, and FS3}, have magnetic fields {preferentially oriented}
in the east-west direction. This {trend} suggests that the field
directions are perpendicular to these filaments {overall at
the angular resolution of our data, {which is too low to separate} FS2 and FS3}. 
A similar {behavior} is observed in the FC2 region. In contrast,
the FN1 and FC1 {filaments exhibit field directions} {which are} mainly parallel 
to {their {major axes}.} Since FN1 and FC1 are less dense {than the other filaments},
this {difference} suggests that magnetic fields are parallel to filaments in less
dense cases but perpendicular in denser ones, {a result that has} been
reported in many previous studies \citep[e.g.,][]{{2016A&A...586A.138P},
{2017ApJ...842...66W}}.
{However}, {magnetic field directions} {{change dramatically}
at the locations of} dense cores.
For example, {the magnetic field seems to} surround the core 
located {{to the south} of FN1, (R.A., decl.) = (18:30:02, +01:15:10).} 
{Field} direction variations are {also clearly
seen} in the central core regions of both the NW and SE subclusters.
A detailed quantitative analysis and discussion {{of} the} relative orientations
of magnetic fields with respect to the filaments {follows} in
Section \ref{sec:discuss}. Comparisons between the magnetic field {on the scales 
resolved in these data and in interferometric observations of} Serpens Main will be
{undertaken} in a separate paper (Tang et al. in preparation).

\begin{figure}
\plotone{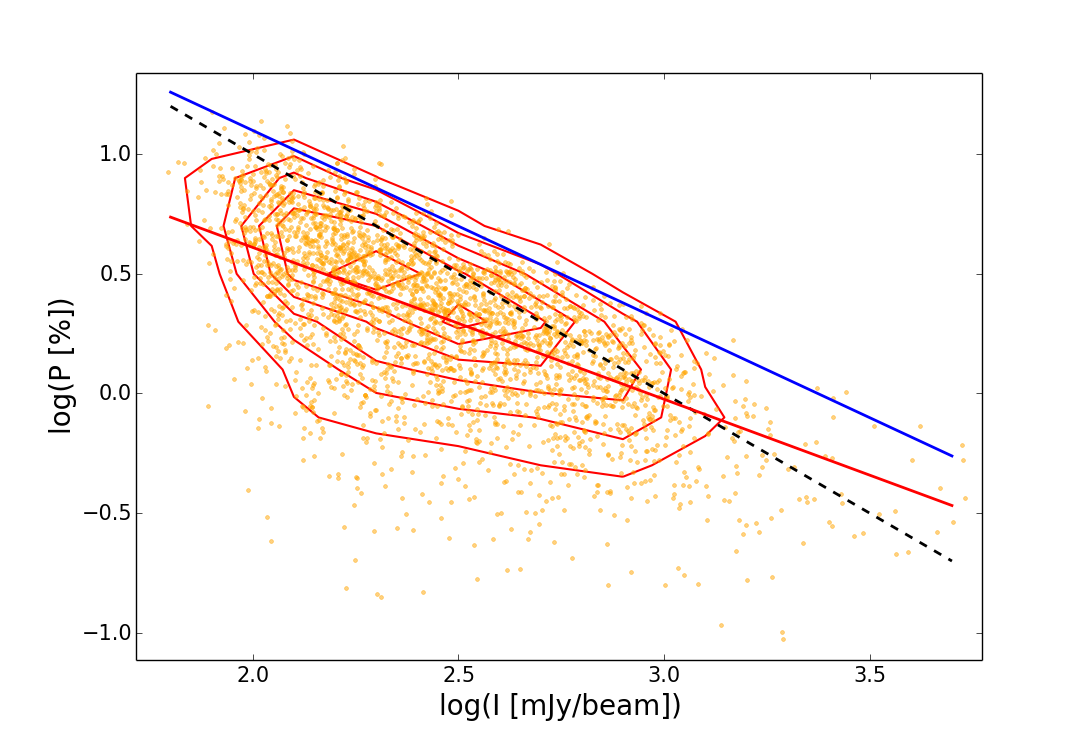}
\caption{{The} relationship between polarization {fraction} and total {intensity}
{in Serpens Main}.
The black dashed, blue solid, and  red solid lines {denote}
slopes of -1, -0.8, and -0.634, respectively.
\label{fig_p_i}}
\end{figure}

\subsection{Polarization fraction as a function of continuum intensity}

The relationship between polarization fraction and column density
(or continuum intensity as a proxy {for} column density)
has been used to show where polarized dust emission originates {{from}} 
in star-forming clouds.
Polarization fractions depend on {both} grain properties and {local physical}
conditions \citep{{2015ARA&A..53..501A}}.  
{In} the case of linear polarization of dust
thermal emission that is {used} in this study, there {must} be a
mechanism {by which} non-spherical dust grains {{are aligned}}.
Such mechanisms
include {anisotropic} radiation, subsonic or supersonic flows,
and/or magnetic fields.
{In the relevant case of} the interstellar medium and molecular
clouds whose dust grains are still sub-micrometer to several
{micrometers in size}, polarization {principally arises from spinning dust grains which are caused to align with respect to the magnetic field by torques from an incident anisotropic radiation field, i.e.}
{both} a magnetic field and {an incident} radiation {field} are needed.

Figure \ref{fig_p_i} presents the relationship between polarization
fraction and continuum intensity {in} Serpens Main {at 850 \um\ on a} {log-log} scale.  {Only} data points that have a SNR
in {total} intensity larger than 30 and {a} polarization
fraction smaller than 15\% have been selected.  The black dashed
line {denotes} a slope of $-1$ {for} reference, {i.e.} $p \propto I^{-1}$.  {The} blue and red solid lines show the trend of
the highest polarization data points 
at {each intensity} ($p
\propto I^{-0.8}$; i.e., the {behaviour of the} upper envelope of the data points) 
and the {result of fitting a power law to the full sample} ($p = 75.6~I^{-0.634}$),
respectively.
Note that the slopes of {both} the {upper-envelope} trend and the fit are
shallower than $-1$. If polarization were limited to the {outer} shell area
of {the} molecular cloud, the relationship between polarization fraction
and intensity would follow the reference slope of $-1$
\citep[e.g.,][]{{2015AJ....149...31J}}.
{The data points in Serpens Main, however,} show a shallow slope 
so polarization is not limited to the {periphery} of the molecular cloud;
some must originate from the dense inner regions of the cloud as well.
This overall trend of polarization fraction decreasing with intensity
is well-known, and often found in molecular clouds
\citep[e.g.,][]{
{2018ApJ...859....4K},{2018ApJ...861...65S},
{2019ApJ...877...43L},{2019ApJ...880...27P}, 
{2019ApJ...876...42W},{2019ApJ...877...88C}}.
This trend is understood {to result from} {weakening} of
dust grain {alignment mechanisms} in high density regions, 
{due to decreasing radiation fields and increasing grain sizes} \citep[e.g.,][]{{2020arXiv201007742H}}.
In addition,
varying magnetic field {directions} along the line of sight depolarize the
signal, {due to integration over multiple magnetic field components.}
Studies of {alternative} polarization mechanisms and dust properties, which are
also interesting and important, have been carried out
\citep[e.g.,][]{{2020A&A...641A..12P},{2020A&A...644A..11L},{Pattle:2020ue}}, {but are unlikely to apply on the size scales that we consider}.

{We note that}
\citet{{2019ApJ...880...27P}} showed that 
as polarization intensities are calculated
{using} the squared sum of Stokes Q and U values,
{resulting in the noise characteristics of} polarization fractions {following a Ricean distribution,}
{an observed} power-law index of $-1$
{of} polarization fraction {as a function of} intensity {may result from the non-detection of polarized emission,}
{raising {the need for}} caution when interpreting {such a}
relationship.
{Our results, however, clearly show a shallower
power-law index, {indicating} real detections.}
Note that in the regime with a power-law index slope shallower 
than $-1$, i.e., {well-detected polarization,}
{both} the Rice distribution and the power-law {approaches}
are consistent with {one another}.

\section{Discussion} \label{sec:discuss}

\subsection{Histogram of relative orientations} \label{subsec:hro}

{The} relationship between magnetic fields and filamentary structures
can be studied {quantitatively using} the histogram of relative orientations
(HRO) technique \citep{2013ApJ...774..128S,{2016A&A...586A.138P}}, 
which compares polarization
directions ($\hat{E}$) {against} {optical depth} gradients ($\nabla \tau$).
{For} the HRO technique, a histogram of relative orientations ($\phi$) 
between $\hat{E}$ and $\nabla \tau$ needs to be produced {at each density to be investigated}.
{In this work, instead} of {using} {optical depth} gradients we {use} intensity gradients for
simplicity, which is reasonable in a small region where 
temperatures and dust properties are not varying {significantly}.
{Note that in the optically thin regime, which is our case,
intensity is proportional to the 
{product of the Planck function for blackbody radiation and the optical depth.}
{Thus,} {a} 15\% variation of temperature, e.g., between
12 K and 14 K, {produces} a 25\% variation in intensity at 850 \um.
In addition, note that the directions of gradients are {analyzed} here, not 
the amplitudes of gradients.}
{Next, {within the HRO the}}
 central area ($A_c$, where $|\phi| < 22.5\degr$)
and the extreme area ($A_e$, where $|\phi| > 67.5\degr$) are calculated.
Finally, the histogram shape parameter ($\xi$) is calculated:
\begin{equation}
\xi = \frac{A_c - A_e}{A_c + A_e}.
\end{equation}
When the relative {orientation} of most data points {is} small, i.e., when
polarization directions are parallel to {optical depth} (here intensity) 
gradients, $\xi$
is positive since {$A_c > A_e$}. In contrast, when polarizations
are perpendicular to intensity gradients, $\xi$ {is} negative.  Since
the magnetic field direction is perpendicular to the polarization
and the filament direction is perpendicular to the intensity gradient,
positive values of the histogram shape parameter $\xi$ also
{imply} that magnetic fields are parallel to filaments.
In the same way, negative $\xi$ values indicate {that} magnetic fields are
perpendicular to filaments.
The uncertainties {on} $\xi$ are 
dominated by the {number} of data points
in {each bin} of the {histogram} and are calculated as \citep{2016A&A...586A.138P}
\begin{equation}
\sigma^2_{\xi} = \frac{4(A_e^2 \sigma_{A_c}^2 + A_c^2 \sigma_{A_e}^2)}{(A_c + A_e)^4}.
\end{equation}
Here, $\sigma_{A_c}$ and $\sigma_{A_e}$ are obtained
following $\sigma_k^2 = h_k (1 - h_k/h_{tot})$, where $h_k$ {corresponds to the number of data points in the central or extreme bin} and $h_{tot}$ {is the total number of data points.}

\begin{figure}
\hspace{-2cm}
\includegraphics[scale=0.25]{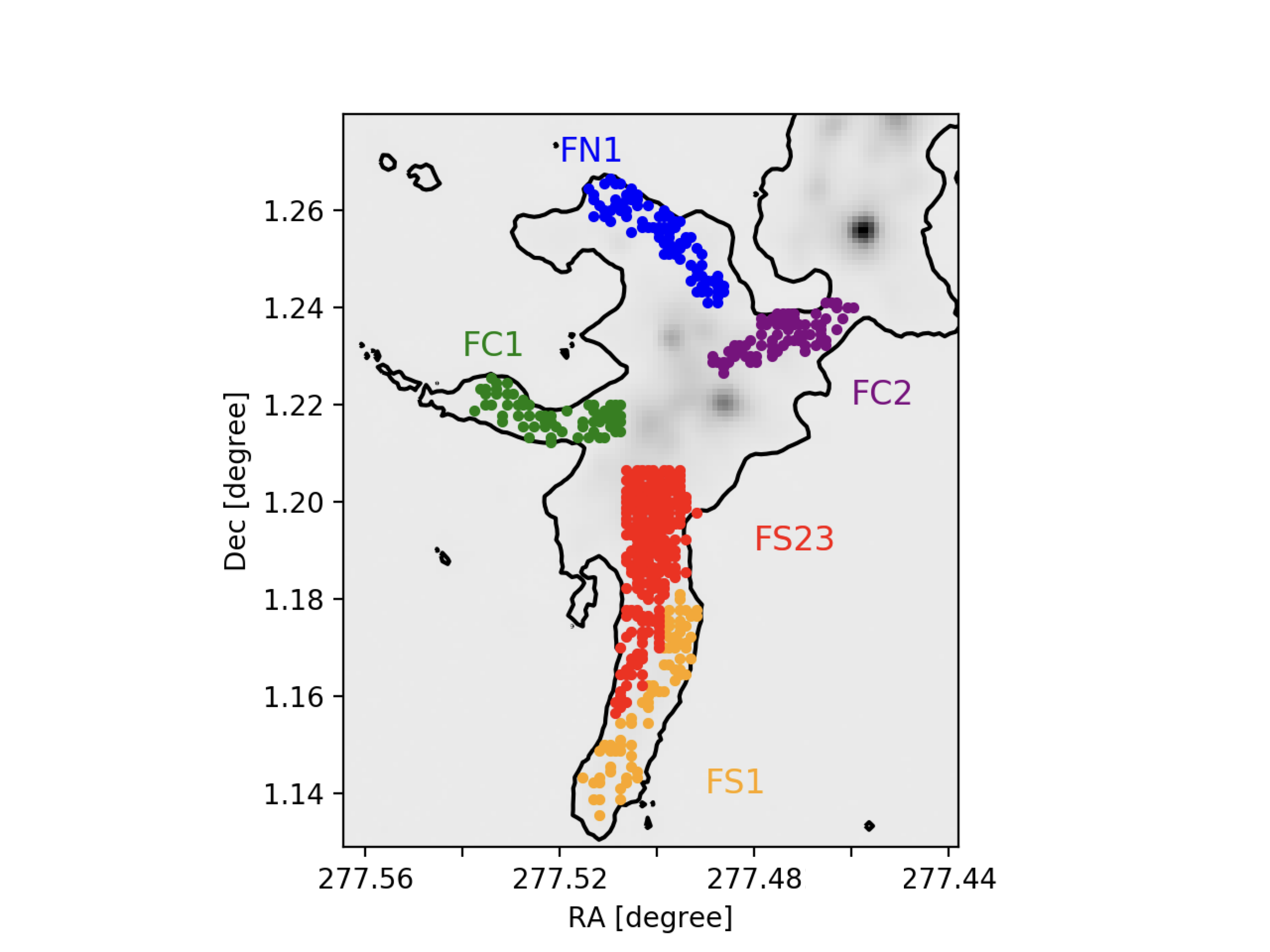}
\hspace{-2cm}
\includegraphics[scale=0.7]{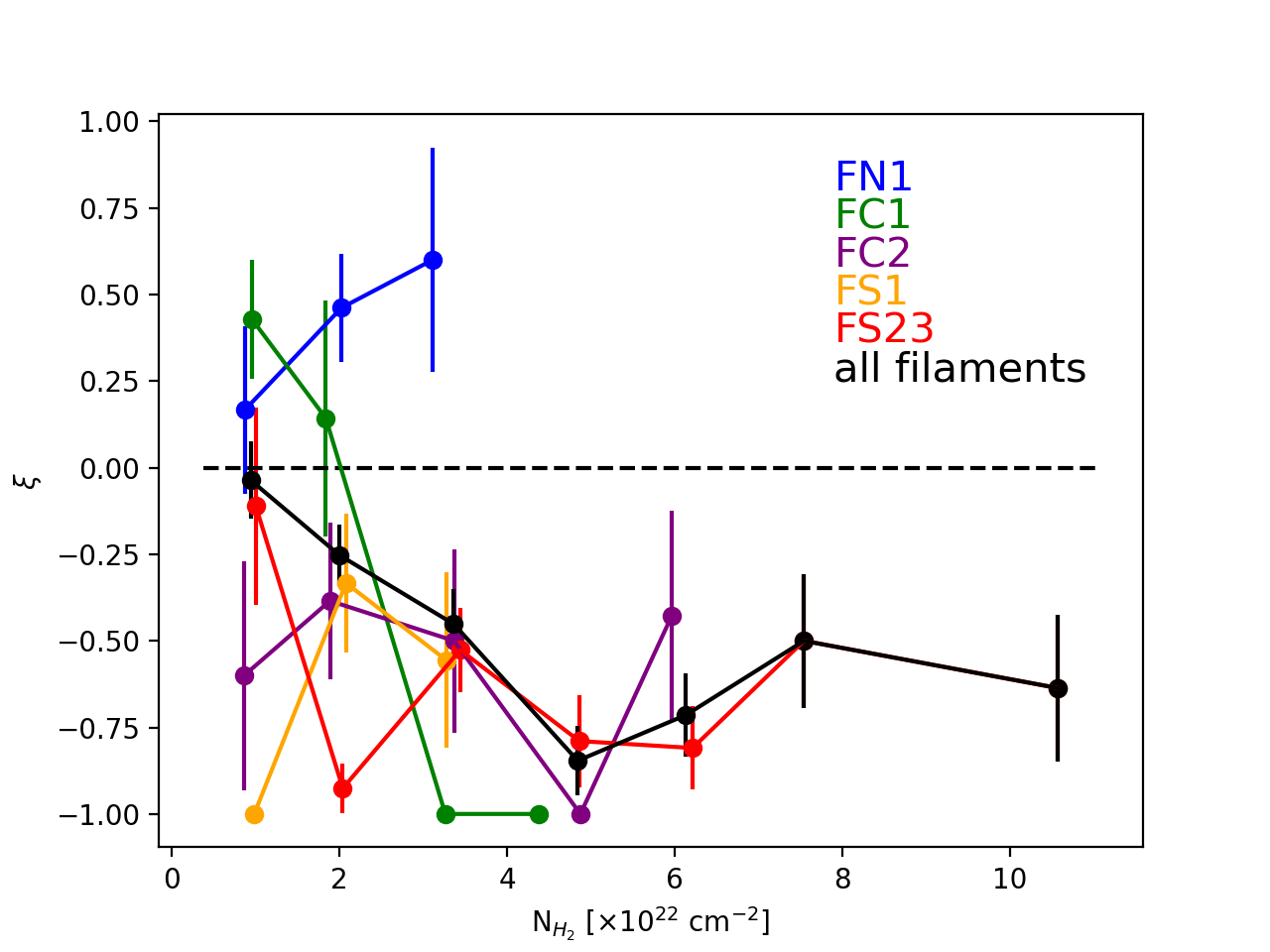}
\caption{
The left panel shows
individual filament regions \citep{{2014ApJ...797...76L}} {used in the HRO analysis}, and
the black contour indicates a level of 0.042 \jybm. {The right panel shows the}
histogram shape parameter ($\xi$) for these filaments, {with the black line indicating the result using all filaments combined}. 
\label{fig_hrofilas}}
\end{figure}

Figure \ref{fig_hrofilas} presents the histogram shape parameters {for
individual filaments as well as for all {filaments combined}.}
The data used for the plot are {gridded to $4\arcsec$ pixels,} with a
SNR in total intensity larger than 3, a SNR in polarized intensity larger
than 2, a polarization angle uncertainty smaller than $15\degr$, and
a polarization fraction smaller than 15\%.
Data points {for} individual filament areas have been selected by {eye}
based on the identifications of \citet{{2014ApJ...797...76L}} and
are presented in the left {panel} of Figure \ref{fig_hrofilas}.
Note that the main trends of {the HRO analysis}
are not changed depending on whether a few segments are included or excluded.
{The column densities ($N_{H_2}$) of molecular hydrogen {used in this analysis}
are calculated from the {850 \um} dust thermal intensities
($I_\nu$):
\begin{equation}
N_{H_2} = \frac{I_\nu}{\kappa_\nu B_\nu(T_d)~\mu_{H_2} m_H \Omega},
\end{equation}
where
$\kappa_\nu$ is the absorption coefficient at 850 \um\ {assuming}
a gas-to-dust mass ratio of 100 (0.02 cm$^2$g$^{-1}$), 
$T_d$ is the dust temperature \citep[12 K,][]{{2014ApJ...797...76L}}, 
$\mu_{H_2}$ is the molecular weight per hydrogen molecule 
\citep[2.8,][]{{Kauffmann:2008jj}},
$m_H$ is the atomic hydrogen mass ($1.67 \times 10^{-24}$ g), and
$\Omega$ is the beam solid angle.}

As shown in Figure \ref{fig_hrofilas} {by the black solid line},
the $\xi$ values {using} all filament data
decrease with column {density}.
The transition from positive to negative $\xi$ is not clear, since
it happens in the first {bin where the} mean intensity 
corresponds to $N_{H_2} \approx 0.93\times 10^{22} \textrm{ cm}^{-2}$.
Therefore, although individual filaments {reveal a range of} properties, which are
addressed below, the overall transition of {the} magnetic field from parallel to
perpendicular {within} the Serpens Main {filaments} tends to occur around this
column density.

Individual filaments {show a range of} relative orientations {for their} associated magnetic
fields, as shown in various colors in Figure \ref{fig_hrofilas}.  First,
all the histogram shape parameters of the filament FN1 are positive
(blue), which means that the magnetic fields in FN1 are parallel
to the filamentary structure. Also, note that {FN1 is the least
dense-filament,} $N_{H_2} < 3.1 \times 10^{22}$ cm$^{-2}$.  {Next}, 
FC1 is the second least dense filament and shows a transition
from positive to negative $\xi$ values (green), which indicates
that magnetic fields are parallel to the filament in its less dense
regions but perpendicular in {the} relatively denser regions.
Note that {the mass-per-length values} of these two filaments 
are subcritical (FN1)
or marginally supercritical (FC1) even without considering magnetic fields, 
and they have no {embedded} YSOs (Table
\ref{tab_filaments}).  {All the} other filaments (FC2, FS1, and FS2/3)
have negative $\xi$ values, {suggesting that their} magnetic fields
are perpendicular to {their {major axes}.}
Filaments FS2 and FS3 are not easy to separate from each other, so
we consider them together. Indeed, {these two filaments} are very similar in density as
the densest filaments among the sample (Table \ref{tab_filaments}).
{{We also note that} there is no significant magnetic field variation over
the region {in response} to the different {observed} filament directions.}

{This trend of {relative} magnetic field orientation changing with density 
has} been reported in
multiple previous studies. For example,
\citet{{2016A&A...586A.138P}} {examined the} relative orientations
between magnetic field directions and column density structures
by applying the HRO technique to {\it Planck} polarimetric data at
353 GHz. {Those authors} found that a transition from parallel to perpendicular
orientations appears between $N_H \sim 0.5 \times 10^{22}$ \cmmt\ and
$\sim 5 \times 10^{22}$ \cmmt, consistent with our result above.

\begin{figure}
\plotone{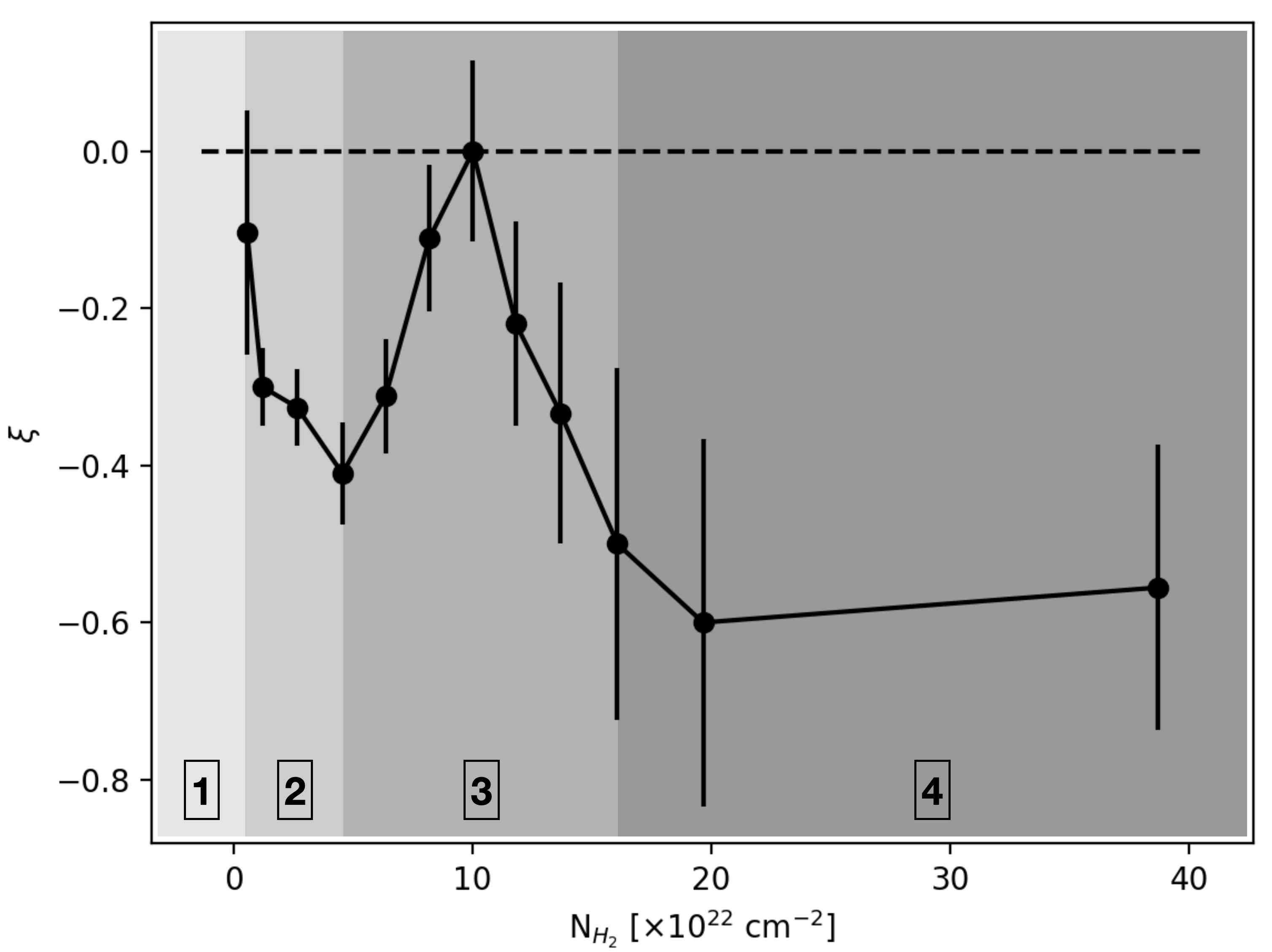}
\caption{Histogram shape parameters ($\xi$) of 
the Serpens Main molecular cloud {including both the filament and
the core regions}.
Note that the least dense regions have {an almost} positive value of $\xi$, and
there is a bump 
{at moderate column density}: $\xi$
becoming positive and decreasing again. Refer to the text for
discussion. The region numbers correspond to those in Figure \ref{fig_diagram}
and in the text. 
\label{fig_hroall}}
\end{figure}

We {further} investigated the magnetic field orientation 
{{by applying} the}
HRO analysis {{to all of} the dense region observations}.
Figure \ref{fig_hroall} shows the histogram shape parameters ($\xi$) 
calculated from the whole of Serpens Main including
the central subcluster areas as well as the filamentary structures.
In the figure, the regime with a column density less than
$N_{H_2} \sim 4.6 \times 10^{22}$ \cmmt\ is mainly {that} of the filamentary structures,
in which the $\xi$ values decrease with density.
Between $N_{H_2} \sim 4.6 \times 10^{22}$ and $\sim 10 \times
10^{22}$ \cmmt, however, the $\xi$ values increase up to zero.
This increase suggests that the magnetic fields are trending 
parallel to 
{elongated structures}
around {a} column density of $N_{H_2} \sim 10 \times
10^{22}$ \cmmt. In other words,
magnetic fields become perpendicular to density gradients
(where {elongated}
spines are defined as being orthogonal to 
the direction of a density gradient). 
The transition can be understood
{as a result of the local} density gradients 
becoming more significant than
the initial density gradients across filaments, presumably {
arising from} mass accumulation of neutral species
crossing magnetic field lines \citep[ambipolar diffusion; e.g.,][]{{1979ApJ...228..475M},{Hennebelle:2019cc}}.  
Also, this regime corresponds to the area where filaments {transition} into
the central hub {while maintaining} {their} initial perpendicular fields.
{With increasing density, the} $\xi$ values again decrease and
get back to the overall decreasing trend around $N_{H_2} \sim 16 \times
10^{22}$ \cmmt, which indicates that magnetic fields are {again} parallel
to density gradients (i.e., perpendicular to ``filaments'').  This {
behavior} can be understood as magnetic field lines {now being dragged
along {by gravitational collapse} at these high column densities.}

\begin{figure}
\plotone{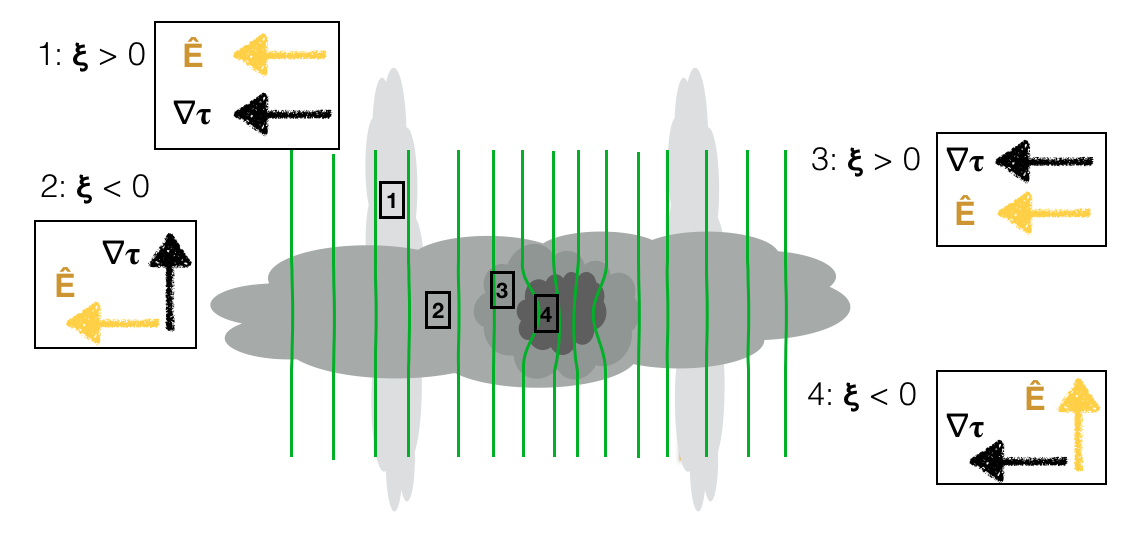}
\caption{Schematic diagram showing four regions with different
relationships between {the} magnetic field and density gradient.
Note that the green lines are magnetic field lines and the darker
gray colors indicate denser regions.
The yellow vectors are polarization directions that are perpendicular
to the magnetic field lines, and the black vectors are 
optical depth gradients that correspond to density gradients. 
The four regions are
(1) less dense filaments, (2) dense filaments, (3) {core-forming} regions,
and (4) {regions in which the field is dragged in}.
\label{fig_diagram}}
\end{figure}

The HRO analysis toward the Serpens Main
molecular cloud nicely shows the three transitional
boundaries in column {density} for relative orientations of magnetic
fields {with respect to} the {physical} structures of filaments and cores. 
Figure \ref{fig_diagram} summarizes the four column density regimes {within}
the Serpens Main molecular cloud {as} a schematic diagram.
First, magnetic fields are parallel to the {least} dense filaments, particularly
where $N_{H_2} \lesssim 0.93\times 10^{22}$ \cmmt\ (region 1). This 
column density threshold corresponds to $n({H_2}) \approx 1.0 \times 10^5$ \cmmth, 
assuming a 0.03 pc {filament width}. 
In contrast,
denser filaments {ranging from} $0.93\times 10^{22} < N_{H_2} \lesssim
4.6 \times 10^{22}$ \cmmt\ have magnetic fields perpendicular to
their filament structure (region 2).
Where {the} column density becomes larger than $N_{H_2} \sim 4.6 \times 10^{22}$ \cmmt,
cores {form and} a significant density gradient along a filament {appears}
(region 3). This column density corresponds to
$n(H_2) \approx 5.0 \times 10^5$ \cmmth,
assuming cores of about 0.03 pc {in size}.
{{Note that} assuming a filament width of 0.03 pc, the critical mass per length 
$M_{L,crit} = 20~M_\sun~pc^{-1}$ corresponds to $N_{H_2} = 3.0 \times 10^{22}$
\cmmt.  
{Considering the presence of magnetic fields, it is interesting that the column density associated with cores in Serpens Main is larger than this value.}
}
Finally, at column densities {above} $N_{H_2} \sim 16 \times 10^{22}$
\cmmt, 
magnetic fields are also dragged in {
by frequent collisions between charged and neutral species during
the dynamical contraction,}
resulting in magnetic fields that are {again} parallel to
the density gradient (region 4). When assuming the same {size scale for cores as previously} (0.03 pc),
{this column density threshold} corresponds to $n(H_2) \approx 18 \times 10^5$ \cmmth\
{or higher.}

In addition to the transition of magnetic field relative orientations
between {less dense and more dense} filamentary structures, {this} further
transition occurring at a higher density, {which we recover,} has recently been reported
by multiple {other} studies. \citet{{Pillai:2020df}} showed that
magnetic fields become parallel to filaments at {a} visual extinction
$A_v \approx 20-30$ magnitude in the Serpens South cloud.  {That} extinction
is equivalent to $N_H \approx 4.5 \times 10^{22} - 6.6 \times 10^{22}$ \cmmt\
($N_{H_2} \approx 1.6 \times 10^{22} - 2.4 \times 10^{22}$ \cmmt), 
based on the relation
between visual extinction and column density
\citep{{2009arXiv0903.2057G}}.
Although {their} column density {threshold differs from ours} by a factor of 2, 
the transition {likely corresponds with} our boundary between regions 2 and 3 {(Figure \ref{fig_diagram})}.  
Also, \citet{{2020arXiv201101555W}}
found that the filament and hub system G33.92+0.11 has a magnetic
field parallel to the gravity {vector} and the gas motion {within} the central hub
region.  Since a gravity direction is comparable to a density
gradient, it {appears that} the hub region {corresponds to region 4
of our scenario.}


\subsection{Field strength and star formation}

Magnetic field strength, {specifically} {the} component projected
on the plane of the sky, can be {estimated} from linearly polarized 
dust thermal emission {using} the Davis-Chandrasekhar-Fermi
(DCF) method \citep{{1951PhRv...81..890D},{1953ApJ...118..113C}}.
The DCF method estimates a field strength at a given {gas} density and
turbulence {from} {the scatter of the magnetic field orientation
around a mean value.}
{This} {scatter} decreases in a stronger field {such that}:
\begin{equation}
B_{POS} = Q \sqrt{4\pi \rho}\frac{\delta V}{\delta \phi} \approx
9.3 \sqrt{n(H_2)} \frac{\Delta V}{\delta \phi},
\end{equation}
where Q is a {a factor of order unity, here taken to be 0.5}, $\rho$ is the density,
$\delta V$ is the turbulent velocity dispersion, and
$\delta \phi$ is the {magnetic} field position-angle dispersion.
In the rightmost expression, 
$n(H_2)$ is a number density of hydrogen molecules in cm$^{-3}$,
$\Delta V$ is a FWHM linewidth in km s$^{-1}$,
and $\delta \phi$ is the {magnetic} field position-angle dispersion in degrees,
which results in a field strength {measured} in $\mu$G
\citep[e.g.,][]{{2001ApJ...546..980O},{2004ApJ...600..279C}}.

\begin{figure}
\includegraphics[scale=0.67]{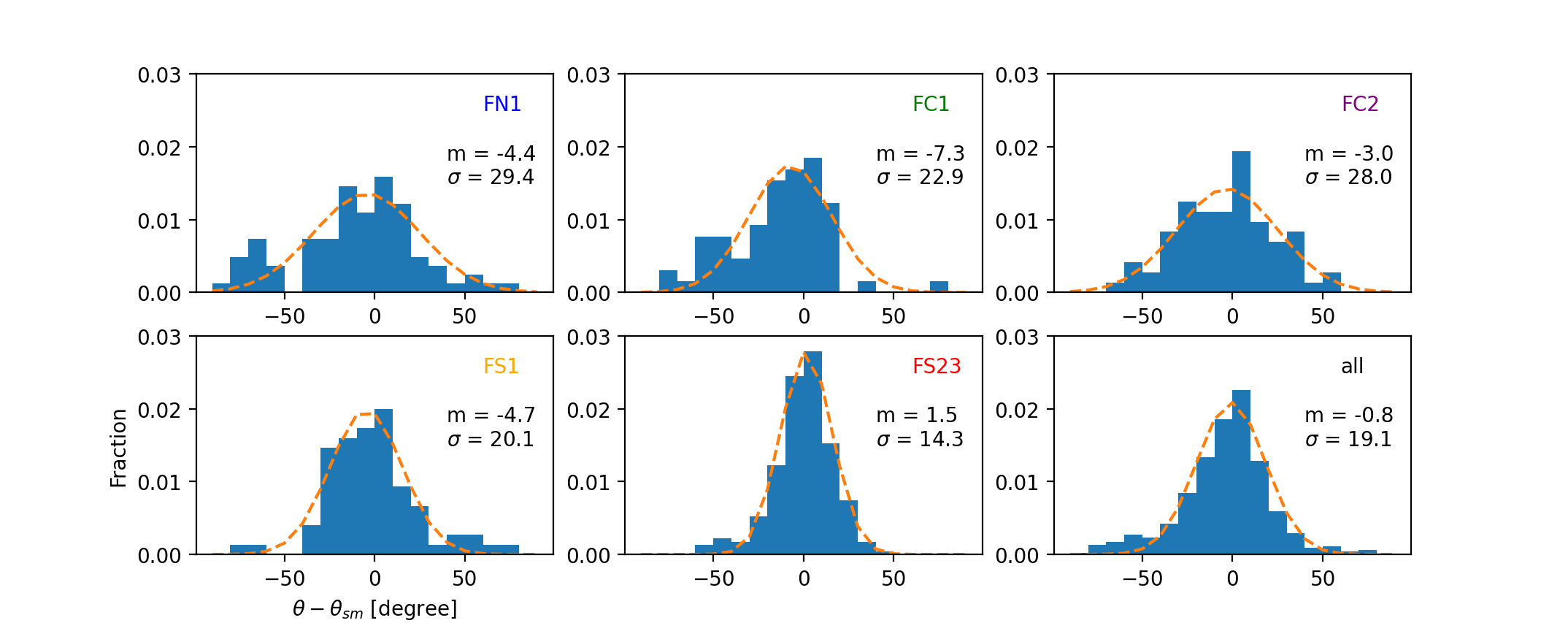}
\caption{
Magnetic field dispersion {for individual filaments} and all filaments {combined}.
{In each panel,} the horizontal axis is position-angle difference in degrees between
the magnetic field direction {of the} non-smoothed and smoothed data.
The orange dashed lines are Gaussian functions {fitted to} the histograms,
whose mean and sigma values are listed in the panels.
\label{fig_dispers}}
\end{figure}

{Table \ref{tab_fields} lists our estimates for} the magnetic field strength {within each of} the filaments
in Serpens Main, using the DCF technique.
The number densities are calculated from
the mean column density of the individual filaments (Fig. \ref{fig_hrofilas})
{divided by}
the filament widths {listed in Table \ref{tab_filaments}}
\citep{{2014ApJ...797...76L}}, {assuming no additional line of sight modification}.
Non-thermal velocity dispersions are adopted from
spectral-line observations of \nthp\ {made} by
\citet{{2014ApJ...797...76L}}, {also} as shown in Table \ref{tab_filaments}.
The {model of ordered} background magnetic field morphology that is {required} {in order to accurately calculate}
a field dispersion is obtained from the
Stokes Q and U maps smoothed {to a resolution of $28 \arcsec$, which is equivalent to the {$2\times$} larger beam}.
{This} $28\arcsec$ {resolution} corresponds to 0.056 pc at the target distance, which
is reasonable for the background {ordered} fields of filamentary structures with a
width $\approx 0.03-0.05$ pc.
The field dispersions {we measure} are distributed from $14.3\degr$ to $29.4\degr$
and {are} plotted in Figure \ref{fig_dispers}.
These values provide magnetic field strengths on the plane of the sky
{ranging from} $64 - 300~\mu$G.
{When estimating} the uncertainty {on} the {magnetic} field strength
we assume a 50\% error {for} the densities and {use} the fitting error {for}
the field position-angle dispersion. {Combined, this results} in 25--42\% uncertainty {in magnetic field strength}.

We also {estimate} the mass-to-flux ratios ($\lambda$) with respect to
their critical values, which {measure} whether the structures are supported
by magnetic fields or if they are expected to gravitationally contract:
\begin{equation}
\lambda = \frac{(M/\Phi)_{obs}}{(M/\Phi)_{crit}} 
= 7.6 \times 10^{-21} \frac{N(H_2)}{B},
\end{equation}
where $N(H_2)$ is a column density of hydrogen molecules in {units} of
$cm^{-2}$ and $B$ is a magnetic field strength in $\mu$G 
\citep[e.g.,][]{{2004ApJ...600..279C}}. 
Regions with $\lambda > 1$ can gravitationally contract.
The uncertainties {on} $\lambda$ are calculated as $\sigma_\lambda
= \sqrt{(\sigma_N/N(H_2))^2 + (\sigma_B/B)^2}$, so
they are in {the} range of 56--65\% when 
{adopting the uncertainties used in the previous paragraph.}
{{Note that this measure of uncertainty provides only a rough estimate, as we assume Gaussian distributions for the uncertainties on column density and the magnetic field strength, which may in fact not be independent of each other.}}
Table \ref{tab_fields} shows the $\lambda$ values {for each filament}.
Note that the values are based on
only the projected magnetic field component. 
Therefore, depending on the {{three dimensional (3D)}} orientation of the magnetic fields,
$\lambda$ {may be} smaller.
We also compare the magnetic field {energy density} ($P_B$) and the
turbulence {energy density} ($P_{turb}$) of individual filaments, which are {respectively} calculated
as
\begin{eqnarray}
P_B &=& \frac{B^2}{8\pi}, \\
P_{turb} &=& \frac{3}{2} \rho \sigma_{nt}^2.
\end{eqnarray}
Here 
$\sigma_{nt}$ {is} non-thermal velocity dispersion. 
{Finally,} the last column of Table \ref{tab_fields} presents the
ratio of mass-per-length to the critical value {for a} cylindrical structure
ignoring magnetic fields, {as} reported by \citet{{2014ApJ...797...76L}}
(Table \ref{tab_filaments}).
{Values} greater than 1 indicate that such a structure can gravitationally
contract.

From these measurements, we classify the
filaments into three groups: \{FN1, FC1\}, \{FC2, FS1\}, and \{FS2/3\}.
In FN1 and FC1, the magnetic and turbulent {energy densities} are comparable to each other.
Note that the $\lambda$ and the magnetic field {energy density} are calculated
{from partial components} of the fields, {specifically} {the component projected} on the plane of the sky.
{When the 3D orientation of the structure is considered,}
the total magnetic field {may} be significantly larger. 
For example, if the {3D} magnetic field strength is larger than the
plane-of-the-sky component by a factor of 2,
$\lambda$ and the magnetic field {energy density} become {halved and quadrupled,}
respectively.
{The FN1 and FC1 filaments both} have $\lambda$ and line mass ratios
either smaller than or close to 1,
which suggests that these filaments may not be able to contract
gravitationally.
It is noteworthy that {these filaments} have no star formation
activity (no YSOs) and {that} they have large velocity gradients, {of} 3--5 
km s$^{-1}$ pc$^{-1}$ (Table \ref{tab_filaments}).

FC2 and FS1 also have magnetic field and turbulence {energy densities} comparable to
{one another}. 
These filaments, however, have both $\lambda$ and mass ratios 
larger than 2, which suggests that they are likely to gravitationally contract.
Indeed, there are YSOs found in FS1 \citep{{2014ApJ...797...76L}}.
In the FC2 filament, there are no YSOs found, but it seems to be {gravitationally unstable}. 
{{If FC2 is in fact not gravitationally unstable, our results suggest that significant magnetic support would be required in order to stabilize the structure.  For such support to exist, based on the plane-of-sky values of $\lambda$ that we measure, the filament would require a significant fraction of its magnetic field to be directed along the line of sight.}}

Lastly, FS2/3, which is the combined region of FS2 and FS3, has the
strongest magnetic field among the filaments.  {The} magnetic
field {energy density} is {also} twice {as large as} the turbulence {energy density}.
Therefore, magnetic
fields dominate {over} turbulence in this {combined} structure.  
Even with the relatively strong magnetic field {compared} to turbulence,
the filaments have $\lambda > 1$ and evidence for YSOs,
indicating that the {regions remain}  gravitationally unstable.

\begin{deluxetable*}{ccccccccc}
\tablecaption{Magnetic field strength {measures for} filaments \label{tab_fields}}
\tablehead{
\colhead{Filament} & 
\colhead{$N_{H_2}$} & 
\colhead{$n(H_2)$} &
\colhead{$\delta \phi$} &
\colhead{$B_{POS}$} &
\colhead{$\lambda$} &
\colhead{P$_{B_{POS}}$} & 
\colhead{P$_{turb}$} & 
\colhead{M$_L$/M$_{L,crit}$} \\
\colhead{} & 
\colhead{[$\times 10^{22}~cm^{-2}$]} & 
\colhead{[$\times 10^{5}~cm^{-3}$]} & 
\colhead{[$\degr$]} &
\colhead{[$\mu$G]} &
\colhead{} &
\colhead{[$\times 10^{-10}$ erg cm$^{-3}$]} & 
\colhead{[$\times 10^{-10}$ erg cm$^{-3}$]} & 
\colhead{}}
\startdata
FN1 & 2.6 & 2.8 & 29.4  & $83\pm35$ & $2.4\pm1.6$ & 2.73  & 6.21 & 0.76 \\
FC1 & 2.2 & 1.8 & 22.9 & $97\pm32$ & $1.7\pm1.0$ & 3.74 & 5.14 & 1.4  \\
FC2 & 4.4 & 2.9 & 28.0 & $84\pm26$ & $4.0\pm2.4$ & 2.78 & 5.72 & 2.4  \\
FS1 & 3.0 & 2.4  & 20.1 & $64\pm17$ & $3.5\pm2.0$ & 1.65 & 1.75 & 2.6  \\ 
FS2/3 & 6.5 & 7.0 & 14.3 & $300\pm75$ & $1.7\pm0.95$ & 34.7 & 18.6 & 3.9  \\
\enddata
\tablecomments{Only the component of the magnetic field on
the plane of the sky has been considered for the 
$\lambda$ and $P_{B_{POS}}$ {measures}.}
\end{deluxetable*}

\section{Conclusions} \label{sec:final}

We have studied {the physical conditions associated with the} magnetic field {within} the Serpens Main
molecular cloud {using JCMT POL-2 850 $\mu$m polarimetric data, with a $14\arcsec$ resolution,
 taken as part of the BISTRO {survey}}.
Serpens Main {consists of two subclusters {containing in total of}
six filamentary structures (FN1, FC1/2, FS1/2/3) } {encompassing a variety of}
physical properties.

The most 
{common} direction of {the} linear polarization in Serpens Main, on $\sim 6000$ au scales, {has} P.A. = $4.5\degr$  which
suggests that the overall magnetic fields are {East-West 
on the plane of sky.}
The {submillimeter} polarization fraction {across}  Serpens Main decreases with the
total intensity as $p \propto I^{-0.634}$.
A power-law index shallower than $-1$ implies that
{the polarized emission} is not limited to the surface of the {molecular cloud}.
 {{The observed}} polarization fraction decrease
can be understood {{as being the result of} weakening} alignment mechanisms
{and/or} magnetic field variations as a function of increasing density
{within} the cloud.

{Using} the HRO technique,
we {find} that the magnetic fields are predominantly perpendicular to
the filament structures {for} all filaments except FN1 and FC1, which 
are the least dense and have no {YSOs}.
The field orientation {transitions} from parallel ($\xi > 0$) 
to perpendicular ($\xi < 0$) {with respect}
to the filaments {when}
$N_{H_2} \approx 0.93\times10^{22}$ \cmmt.
We extend the typical HRO analysis to explore the magnetic 
field {within} the entire cloud and {find} that 
$\xi$ again increases 
at $N_{H_2} \approx 4.6\times10^{22}$ \cmmt, which
can be understood {as the result of} a density gradient along 
{elongated structures}
{becoming} significant (e.g., via core formation).
Furthermore, $\xi$ {again becomes} negative at
$N_{H_2} \approx 16\times10^{22}$ \cmmt, which
is interpreted as {resulting from the}  magnetic field {being dragged along
with increasing density.}

We estimate magnetic field strengths {in} the filaments {ranging from} $B_{POS} = 60-300 ~\mu$G, using the DCF method.  The magnetic
field {energy density} is comparable to or larger than the {turbulent energy}
in all filaments.
The starless filaments FN1 and FC1 {might} be supported by 
either the magnetic field or turbulence,
while the other filaments appear to be gravitationally unstable. 
{{The combined filaments} FS2/3, with the strongest magnetic field, also {seem} to be
gravitationally unstable.}

\acknowledgments
We are grateful to the anonymous referee for helpful comments.
W.K. is supported by the New Faculty Startup Fund from Seoul
National University.
{K.P. is a Royal Society University Research Fellow, supported by grant number URF\textbackslash R1\textbackslash 211322.}
{D.J.\ is supported by NRC Canada and by an NSERC Discovery Grant.}
C.L.H.H. acknowledges support from the NAOJ Fellowship and JSPS
KAKENHI grants 18K13586 and 20K14527.
M.T. is supported by JSPS KAKENHI grant Nos.18H05442,15H02063,and 22000005.
K.Q. is partially supported by National Key R\&D Program of China
No. 2017YFA0402600, and acknowledges the National Natural Science
Foundation of China (NSFC) grant U1731237.
C.W.L. is supported by the Basic Science Research Program through the NRF funded
by the Ministry of Education, Science and Technology (NRF-2019R1A2C1010851).
C.E. acknowledges the financial support from grant RJF/2020/000071 as a part of Ramanujan Fellowship awarded by Science and Engineering Research Board (SERB), Department of Science and Technology (DST), Govt. of India.
L.F. acknowledges the support of the Ministry of Science and Technology of Taiwan under grant MOST107-2119-M-001-031-MY3 and Academia Sinica under grant AS-IA-106-M03.

This study is part of the JCMT Large Program BISTRO observed under
the project code M16AL004.
The James Clerk Maxwell Telescope is operated by the East Asian
Observatory on behalf of The National Astronomical Observatory of
Japan; Academia Sinica Institute of Astronomy and Astrophysics; the
Korea Astronomy and Space Science Institute; Center for Astronomical
Mega-Science (as well as the National Key R\&D Program of China with
No. 2017YFA0402700). Additional funding support is provided by the
Science and Technology Facilities Council of the United Kingdom and
participating universities and organizations in the United Kingdom
and Canada.
Additional funds for the construction of SCUBA-2 were provided
by the Canada Foundation for Innovation.
The authors wish to recognize and acknowledge the very significant
cultural role and reverence that the summit of Maunakea has
always had within the indigenous Hawaiian community. We are most
fortunate to have the opportunity to conduct observations from this
mountain.

%

\vspace{5mm}
\facilities{JCMT}

\bibliographystyle{apj}
\bibliography{serpens}



\end{CJK*}
\end{document}